\documentclass{article}
\author{Edward Kapu\'scik\\Professor Emeritus\\University of \L\'od\'z and Institute of Nuclear Physics PAS, Krak\'ow\\ Poland}
\title{On a  Fatal Error in Tachyonic Physics} 
\begin{document}
\maketitle

\begin{abstract}
A fatal error in the famous paper on tachyons by Gerald Feinberg is pointed out. The correct expressions for energy and momentum of tachyons are derived.
\end{abstract}

It is well-known that in Special Theory of Relativity the energy $E$ and momentum $\vec{P}$ of a point particle with mass $M$ are given by\cite{1}
$$
E=\frac{Mc^2}{\sqrt{1-\frac{\vec{v}^2}{c^2}}}
\eqno(1)
$$
and
$$
\vec{P}=\frac{M\vec{v}}{\sqrt{1-\frac{\vec{v}^2}{c^2}}},
\eqno(2)
$$
where $\vec{v}$ is the velocity of the particle. The presence of the square root in (1) and (2) implies that $|\vec{v}|<c$.

In order to overcome the limitation for velocities of motion, Gerald Feinberg in 1967  \cite{2} proposed to replace the mass $M$ in (1) and (2) by an imaginary quantity $i\mu$. This allows to convert the expressions for energy and momentum  into
$$
E=\frac{\mu c^2}{\sqrt{\frac{\vec{v}^2}{c^2}-1}},
\eqno(3)
$$
$$
\vec{P}=\frac{\mu\vec{v}}{\sqrt{\frac{\vec{v}^2}{c^2}-1}},
\eqno(4)
$$
where $\vec{v}$ is the tachyon velocity and $\mu$ is the so-called metamass of the tachyon. Clearly, this time $|\vec{v}|>c$. It must be however stressed that both epressions (3) and (4) were not derived from some first principles but just proposed \textit{ad hoc} in order to allow the tachyons to move faster than light.

As a consequence of these formulas G. Feinberg pointed out that the energy-momentum four-vector $(E, \vec{P})$ of tachyons is a spacelike fourvector since
$$
E^2-c^2\vec{P}^2=-\mu^2c^4<0
\eqno(5)
$$
and therefore
$$
c|\vec{P}|>E.
\eqno(6)
$$
From the Lorentz transformation for the energy-momentum four-vector of the form
$$
E\rightarrow E'=\gamma\left(E-R\vec{P}\cdot\vec{V}\right),
\eqno(7)
$$
$$
\vec{P}\rightarrow \vec{P}'=R\vec{P}+(\gamma-1)\frac{R\vec{P}\cdot\vec{V}}{\vec{V}^2}\vec{V}-\gamma\frac{E}{c^2}\vec{V},
\eqno(8)
$$
where $\gamma$ is the famous Lorentz factor given by
$$
\gamma=\left(1-\frac{\vec{V}^2}{c^2}\right)^{-1/2},
\eqno(9)
$$
with $\vec{V}$ the relative velocity of two observers and $R$  the rotation of space axes used by the observers,  G. Feinberg pointed out that in some reference frames the energy of tachyons becomes to be negative because in view of (6) for some velocities $\vec{V}$ the expression in the parenthesis in (7) certainly is negative.

The question however arises: how the evidently positive quantity, given in all reference frames by (3), may become to be negative?

The resolution of this paradox consists in the observation that the proposed quantities (3) and (4), as a matter of fact, \textbf{are not components of a fourvector} and therefore under Lorentz transformations they do not transform according to (7) and (8). In fact, taking into account that quantities (3) and (4), as a matter of fact, \textbf{are not components of a fourvector} and therefore under Lorentz transformations they do not transform according to (7) and (8). In fact, taking into account that
$$
E'\equiv E(\vec{v}')
\eqno(10)
$$
and
$$
\vec{P}'\equiv\vec{P}(\vec{v}')
\eqno(11)
$$
with
$$
\vec{v}'=\frac{R\vec{v}+(\gamma-1)\frac{\vec{V}\cdot R\vec{v}}{\vec{V}^2} \ \vec{V}-\gamma\vec{V}}{\gamma\left(1-\frac{\vec{V}\cdot R\vec{v}}{c^2}\right)}
\eqno(12)
$$
from (3) and (4), after some not long calculation, we get
$$
E'=\gamma\left|E-R\vec{P}\cdot\vec{V}\right|
\eqno(13)
$$
and
$$ 
\vec{P}'=sign\left(E-R\vec{P}\cdot\vec{V}\right)\left[R\vec{P}+(\gamma-1)\frac{R\vec{P}\cdot\vec{V}}{\vec{V}^2}\vec{V}-\gamma\frac{E}{c^2}\vec{V}\right].
\eqno(14)
$$
The reason for the appearance of the absolute value in (13) comes from the fact that in the process of calculation we meet the expression
$$
\sqrt{\left(1-\frac{R\vec{v}\cdot\vec{V}}{c^2}\right)^2}=\left|1-\frac{R\vec{v}\cdot\vec{V}}{c^2}\right|.
\eqno(15)
$$     
Clearly, for standard subluminal velocities $\vec{v}$ we have $R\vec{v}\cdot\vec{V}<c^2$. Therefore the expression $1-\frac{R\vec{v}\cdot\vec{V}}{c^2}$ always is positive and the sign of the absolute value in (15) is not needed. As a result we get the standard Lorentz transformation rules (7) and (8). But it is not the case for superluminal speeds $\vec{v}$ for which the expression $1-\frac{R\vec{v}\cdot\vec{V}}{c^2}$ may be both positive and negative. Consequently, the signs of absolute value in (15) and consequently in (13) are necessary. Unfortunately, G. Feinberg did not notice this fact. As a result we must conclude that\textbf{ the Feinberg's expressions (3) and (4) do not provide the energy-momentum fourvector}.

In order to get the correct expressions for energy and momentum of superluminal objects let us observe that the Lorentz transformation rules (7) and (8) together with (10) and (11) lead to the following set of functional equations for energy and momentum
$$
E\left(\frac{R\vec{v}+\left[\gamma(\vec{V})-1\right]\frac{\vec{V}\cdot R\vec{v}}{\vec{V}^2} \ \vec{V}-\gamma(\vec{V})\vec{V}}{\gamma(\vec{V})\left(1-\frac{\vec{V}\cdot R\vec{v}}{c^2}\right)}\right)=\gamma(\vec{V})\left(E(\vec{v})-R\vec{P}(\vec{v})\cdot\vec{V}\right),
\eqno(16)
$$
$$
\vec{P}\left(\frac{R\vec{v}+\left[\gamma(\vec{V})-1\right]\frac{\vec{V}\cdot R\vec{v}}{\vec{V}^2} \ \vec{V}-\gamma(\vec{V})\vec{V}}{\gamma(\vec{V})\left(1-\frac{\vec{V}\cdot R\vec{v}}{c^2}\right)}\right)=\ \ \ \ \ \ \ \ \ \ \ \ \ \ \ \ \ \ \ \ \ \ \ \ \ \ \ \ \ \ \ \ \ \ \ \ \ \ \ \ \ 
$$
$$
\ \ \ \ \ \ \ \ \ \ \ \ \ \ \ \ \ \ \ \ \ \ \ \ \ \ \ \ \ \ \ =R\vec{P}(\vec{v})+\left[\gamma(\vec{V})-1\right]\frac{R\vec{P}(\vec{v})\cdot\vec{V}}{\vec{V}^2}\vec{V}-\gamma(\vec{V})\frac{E(\vec{v})}{c^2}\vec{V}.
\eqno(17)
$$

It is not difficult to solve this set of functional equations. For subluminal objects there always exist rest frames and therefore we may put $\vec{v}=0$. Then from  (16) and (17) we get
$$
E\left(-\vec{V}\right)=\gamma(\vec{V})\left(E(\vec{0})-R\vec{P}(\vec{0})\cdot\vec{V}\right),
\eqno(18)
$$
$$
\vec{P}\left(-\vec{V}\right)= R\vec{P}(\vec{0})+\left[\gamma(\vec{V})-1\right]\frac{R\vec{P}(\vec{0})\cdot\vec{V}}{\vec{V}^2}\vec{V}-\gamma(\vec{V})\frac{E(\vec{0})}{c^2}\vec{V}.
\eqno(19)
$$
Requiring that momentum is an odd function of velocity we must put $\vec{P}(\vec{0})=\vec{0}$ and using the Einstein relation $E(\vec{0})\equiv E_0=Mc^2$ we arrive (after changing the sign of $\vec{V}$) to the standard expressions (1) and (2) for energy and momentum of subluminal objects.

For superluminal objects there are no rest frames because the velocities are restricted from below by $c$. The velocities are however not restricted from above and therefore we always may consider the limit of infinite velocities $\vec{v}$ in (16) and (17). In order to execute such a limit let us choose the rotation $R$ in such a way that $R\vec{v}=\lambda \vec{V}$. Then taking the limit $\lim \lambda\rightarrow\infty$ we get
$$
E\left(-\frac{c^2}{\vec{V}^2}\vec{V}\right)=\gamma(\vec{V})\left(E(\infty)-R\vec{P}(\infty)\cdot\vec{V}\right),
\eqno(20)
$$
$$
\vec{P}\left(-\frac{c^2}{\vec{V}^2}\vec{V}\right)  =R\vec{P}(\infty)+\left[\gamma(\vec{V})-1\right]\frac{R\vec{P}(\infty)\cdot\vec{V}}{\vec{V}^2}\vec{V}-\gamma(\vec{V})\frac{E(\infty)}{c^2}\vec{V}.
\eqno(21)
$$
Denoting 
$$
\vec{W}=-\frac{c^2}{\vec{V}^2}\vec{V},
\eqno(22)
$$
we may express the right-hand side in terms of $\vec{W}$. Requiring that the energy should be an even function of velocity while the momentum an odd function of velocity we must put $\vec{P}(\infty)=0$ and finally we get\cite{5}\cite{6}
$$
E(\vec{W})=\frac{E_{\infty}}{\sqrt{1-\frac{c^2}{\vec{W}^2}}},
\eqno(23)
$$
$$
\vec{P}(\vec{W})  =\frac{E_\infty}{\vec{W}^2\sqrt{1-\frac{c^2}{\vec{W}^2}}}\vec{W},
\eqno(24)
$$
where we have introduced the notation $E_\infty=E(\infty)$ for the energy of objects moving with infinite speeds.

It must be stressed that according to (22) the velocity of a tachyon in a given reference frame is always determined in a nonlinear way by the relative velocity of the given reference frame with respect to the reference frame in which the tachyon moves with infinite speed. As a consequence we get the unusual transformation rule for the tachyon velocities in the following form
$$
\vec{W}'=\frac{\gamma\left(1-\frac{\vec{u}\cdot R\vec{W}}{\vec{W}^2}\right)} {1-\frac{\vec{W}^2}{c^2}+\gamma^2\frac{\vec{w}^2}{c^2}\left(1-\frac{\vec{u}\cdot R\vec{W}}{\vec{W}^2}\right)^2}
\left[R\vec{W}+(\gamma-1)\frac{\vec{u}\cdot R\vec{W}}{\vec{W}^2}\vec{u}-\gamma\frac{\vec{W}^2}{c^2}\vec{u}\right]
\eqno(25).
$$

The above formulas imply several important conclusions. First, they rigorously follow from the general principles of special theory of relativity and therefore are well justified. Second, the energy-momentum fourvector given by (23) and (24), contrary to the Feinberg energy-momentum "fourvector", is a timelike fourvector because
$$
E^2-c^2\vec{P}^2=E^2_\infty>0.
\eqno(26)
$$
This fact eliminates all troubles in constructing quantum field theory of tachyons \cite{4}. In particular, tachyons may exist with arbitrary spins contrary to the Feinberg tachyon which can be only spinless. Third, for infinite velocities momentum of tachyons vanishes while their energy remains finite. Fourth, from the expressions (23) and (24) it follows that the tachyon velocity is given by
$$
\vec{W}=\frac{E}{\vec{P}^2}\\ \vec{P}.
\eqno(27)
$$ 
which must be confronted with the expression for the velocity of the subluminal particles
$$
\vec{v}=\frac{\vec{P}}{E}
\eqno(28)
$$
which follow from (1) and (2). Eq.(27) supports the Feinberg argumentation that in order to determine the tachyon velocity it is sufficient to measure momentum and energy of tachyons. However, for the same values of momentum and energy, applying the relation (25), we may also obtain subluminal velocities. To resolve the dilemma whether we have to do with subluminal or with superluminal velocities we must use the time of flight method.

\end{document}